\documentclass[sigplan,screen,urlbreakonhyphens=false]{acmart}
\setcopyright{acmcopyright}
\acmPrice{15.00}
\acmDOI{10.1145/3497775.3503692}
\acmYear{2022}
\copyrightyear{2022}
\acmSubmissionID{poplws22cppmain-p45-p}
\acmISBN{978-1-4503-9182-5/22/01}
\acmConference[CPP '22]{Proceedings of the 11th ACM SIGPLAN International Conference on Certified Programs and Proofs}{January 17--18, 2022}{Philadelphia, PA, USA}
\acmBooktitle{Proceedings of the 11th ACM SIGPLAN International Conference on Certified Programs and Proofs (CPP '22), January 17--18, 2022, Philadelphia, PA, USA}
\usepackage{microtype}
\usepackage{balance}
\usepackage{pgf,natbib}
\usepackage[strings]{underscore}
\usepackage{flushend}

\newtheorem{condition}{Condition}

{\bfseries}{\itshape}
\newtheorem{remark}{Remark}
\newtheorem{property}{Property}

\title{A Drag-and-Drop Proof Tactic}

\author{Pablo Donato}
\email{pablo.donato@polytechnique.edu}
\affiliation{
  \institution{École polytechnique}
  \department{LIX}
  \country{France}}

\author{Pierre-Yves Strub}
\email{pierre-yves.strub@polytechnique.edu}
\affiliation{
  \institution{École polytechnique}
  \department{LIX}
  \country{France}}

\author{Benjamin Werner}
\email{benjamin.werner@polytechnique.edu}
\affiliation{
  \institution{École polytechnique}
  \department{LIX}
  \country{France}}

\usepackage{amsmath}
\usepackage{mathtools}
\usepackage{mathpartir}
\usepackage{prftree}
\usepackage{float}
\usepackage{framed}
\usepackage{stmaryrd}
\usepackage{xparse}
\usepackage[font=small]{caption}
\usepackage{multirow}

\NewDocumentCommand{\rnm}{mO{}tr}
  {{\small$
    \IfBooleanT{#3}{\overline}{
      #1
    }^{#2}$}}

\NewDocumentCommand{\R}{oO{}trtd}{
  \expandafter\prftree\expanded{
    \IfBooleanT{#4}{[d]}
    \IfValueT{#1}{[r]{
      \unexpanded{\rnm{#1}}
      [#2]
      \IfBooleanT{#3}{r}
    }}
  }
}

\newcommand{\rnmsf}[1]{\rnm{\mathsf{#1}}}

\newcommand{\rever}{^*}
\newcommand{\hole}{\square}

\newcommand{\substi}[2]{#1\setminus#2}
\newcommand{\substs}[2]{#1\left[#2\right]}
\newcommand{\subst}[3]{\substs{#1}{\substi{#2}{#3}}}

\newcommand{\pair}[2]{\left\langle #1, #2 \right\rangle}

\newcommand{\select}[1]{\fbox{$#1$}}
\newcommand{\forw}{~~\ast~~}
\newcommand{\back}{~\vdash~}

\newcommand{\inv}{\mbox{\it inv\/}}

\newcommand{\mother}{\mbox{\sf Mother}}
\newcommand{\rich}{\mbox{\sf Rich}}

\DeclareMathOperator{\link}{~@~}

\renewcommand{\implies}{\Rightarrow}
\newcommand{\lto}{\Rightarrow}

\newcommand{\seq}{\vdash}

\DeclareMathOperator{\rew}{~~\rhd~~}
\DeclareMathOperator{\mrew}{~~\rhd^\ast~~}
\DeclareMathOperator{\syneq}{~\equiv~}

\newcommand{\human}{\mbox{\sf Hum}}
\newcommand{\mortal}{\mbox{\sf Mort}}
\newcommand{\socrates}{\mbox{\sf Socr}}

\newcommand{\suc}[1]{#1\!\oplus\! 1}
\begin{document}

\begin{abstract}
  We explore the features of a user interface where formal proofs can
  be built through gestural actions. In particular, we show how proof
  construction steps can be associated to drag-and-drop actions. We
  argue that this can provide quick and intuitive proof construction
  steps. This work builds on theoretical tools coming from deep
  inference. It also resumes and integrates some ideas of the former
  proof-by-pointing project.
\end{abstract}



\maketitle

\section{Introduction}

Most Interactive Theorem Provers allow the user to incrementally construct
formal proofs through an interaction loop. One  progresses
through a sequence of {\em states} corresponding to incomplete proofs. Each
of these states is itself described by a finite set of {\em goals} and
the proof is completed once there are no goals left.
{\bf}
From the user's point of view, a goal appears as a sequent, in the
sense coined by Gentzen. In the case of intuitionistic logic that is:
\begin{itemize}
\item One particular proposition $A$ which is {\em to be proved}; we
  designate it as the goal's {\em conclusion},
\item a set of propositions $\Gamma$ corresponding to {\em hypotheses}.
\end{itemize}
On paper, this sequent is written $\Gamma\vdash A$.  The user performs
{\em actions} on one such goal at a time, and the actions transform
the goal, or rather replace the goal by a new set of goals. When this
set is empty, the goal is said to be solved.

The actions performed by the user can be more or less
sophisticated. But, fundamentally, one finds elementary commands which
correspond roughly to the logical rules, generally of natural
deduction. For instance, a goal $\Gamma\vdash A\vee B$
(resp. $\Gamma\vdash A\wedge B$) can be turned into either a goal
$\Gamma\vdash A$ or a goal $\Gamma\vdash B$ (resp. into two goals
$\Gamma\vdash A$ and $\Gamma\vdash B$).

To sum up, during the proof construction process, a state is a set of
sequents. These goals/sequents are modified by {\em commands}, which
allow the user to navigate from the original statement of the theorem
to the state where there are no goals left to be proved.

In the dominant paradigm, these commands are provided by the user in
text form; since Robin Milner and
LCF~\cite{doi:10.1098/rsta.1984.0067}, they are called {\em
  tactics}. Proof files are literally {\em proof-scripts}; that is the
sequence of tactics typed-in by the user.

The present work is a form of continuation of the {\em
  Proof-by-Pointing} (PbP) effort, initiated in the 1990's by Gilles
Kahn, Yves Bertot, Laurent Théry and their group \cite{PbP}. Both works share a
main idea which is to replace the textual tactic commands by {\em
  gestural actions} performed by the user on a graphical user
interface. In both cases, the {\em items} the user performs actions on
are the current goal's conclusion and hypotheses. What is new in our
work is that we allow not only to {\em click} on (subterms of) these
items, but also to {\em move them} in order to drag-and-drop (DnD) one item
onto another. This enriches the language of actions in, we argue, an
intuitive way. We should point out that what is proposed here is not
meant to replace but to complement the proof-by-pointing features. We
thus envision a general {\em proof-by-action} paradigm, which includes
both PbP and DnD features.

In this article, we focus on how drag-and-drop actions implement proof
construction operations corresponding to the core logic; that is how
they deal with logical connectives, quantifiers and equality. We have started to
implement this in a prototype named {\em Actema} (for Active Mathematics)
running through a web HTML5/JavaScript interface.
This possibility to experiment in practice, even though yet on a small scale,
gave valuable feedback for crafting the way DnD actions are to be translated
into proof construction steps in an intuitive and practical way.

The rest of this article is organized as follows. Section \ref{sec:motivations}
explains the motivations behind this work, and section \ref{sec:setting} briefly
outlines its logical setting. Section~\ref{sec:aristote} describes the basic
features of a graphical proof interface based on our principles, and illustrates
them with a famous syllogism from Aristotle. Section \ref{sec:clicks} shows how
it can integrate basic proof-by-pointing capabilities. The next two sections
explain, through further examples, how the drag-and-drop paradigm works; first
for so-called \emph{rewrite} actions involving equalities, then for actions
involving logical connectives and quantifiers. Section~\ref{sec:correctness}
introduces the notions of context and polarity, in order to prove the
correctness of our system. Section \ref{sec:linkages} explains how DnD actions
are specified by the user interactively, through schemas called \emph{linkages}.
Section \ref{sec:action} describes how linkages translate into logical steps, as
well as some properties of this translation. Section \ref{sec:edukera} studies a
proof of a small logical riddle in Actema, highlighting some benefits of our
approach compared to textual systems. We end with a discussion on some related
works in section \ref{sec:related-work}, and then conclude.

\section{Motivations}\label{sec:motivations} Since this work is about changing
the very way the user interacts with an interactive theorem prover, we feel it
is important to make some disclaimers about the aims and the scope of what is
presented here.

From a development point of view, we are still at a very preliminary
stage. Building a real-size proof system integrating the ideas we
present would require an important effort and is still a long term
goal. Some concepts however have emerged, which, we hope allow to
sketch some aspects of the look-and-feel of such a system, and what
some of its advantages could be.

Also, at this stage, we focus on basic proof constructions and on how the
gestural approach can help make them more efficient and more intuitive. Some of
the illustrative examples we give below could probably be dealt with using
advanced proof search tactics, but we believe this does not make them
irrelevant. Rather than (sub)goals to be proved, these examples should be seen
as generic situations often encountered in the course of a proof, which require
small and local transformations to the statements involved.

The idea of interactive theorem provers is that automation and user actions
complement each other, and we here focus on the latter for the time being. The
question of integrating drag-and-drop actions and powerful proof automation
techniques is left for future work.

Finally, precisely because our approach is about giving the user a smoother
control of the proof construction process, we see a possibility for
our work to help making future proof systems more suited for education.

\section{Logical Setting}\label{sec:setting}
Any proof system must implement a given logical formalism. What we
describe here ought to be applied to a wide range of formalisms, but
in this article we focus on the core of intuitionistic first-order logic
with equality (FOL). This allows us to consider sequents where
hypotheses are unordered which, in turn, simplifies the technical
presentations.  We will thus write $\Gamma,A \vdash B$ for a sequent
where $A$ is among the hypotheses.

We use and do not recall the usual definitions of terms and
propositions in first order logic. We assume a first order language
(function and predicate symbols) is given. Provability
is defined over sequents $\Gamma\vdash A$ by the usual logical rules
of natural deduction (NJ) and/or sequent calculus (LJ).

Equality is treated in a common way: $=$ is a binary
predicate symbol written in the usual infix notation, together with the
reflexivity axiom $\forall x.x=x$ and the Leibniz scheme, stating that for any
proposition $A$ one has
$$\forall x.\forall y. x=y\wedge A \implies \subst{A}{x}{y}.$$

\begin{figure*}
 \begin{center}
\fbox {   \includegraphics[width=.9\textwidth]{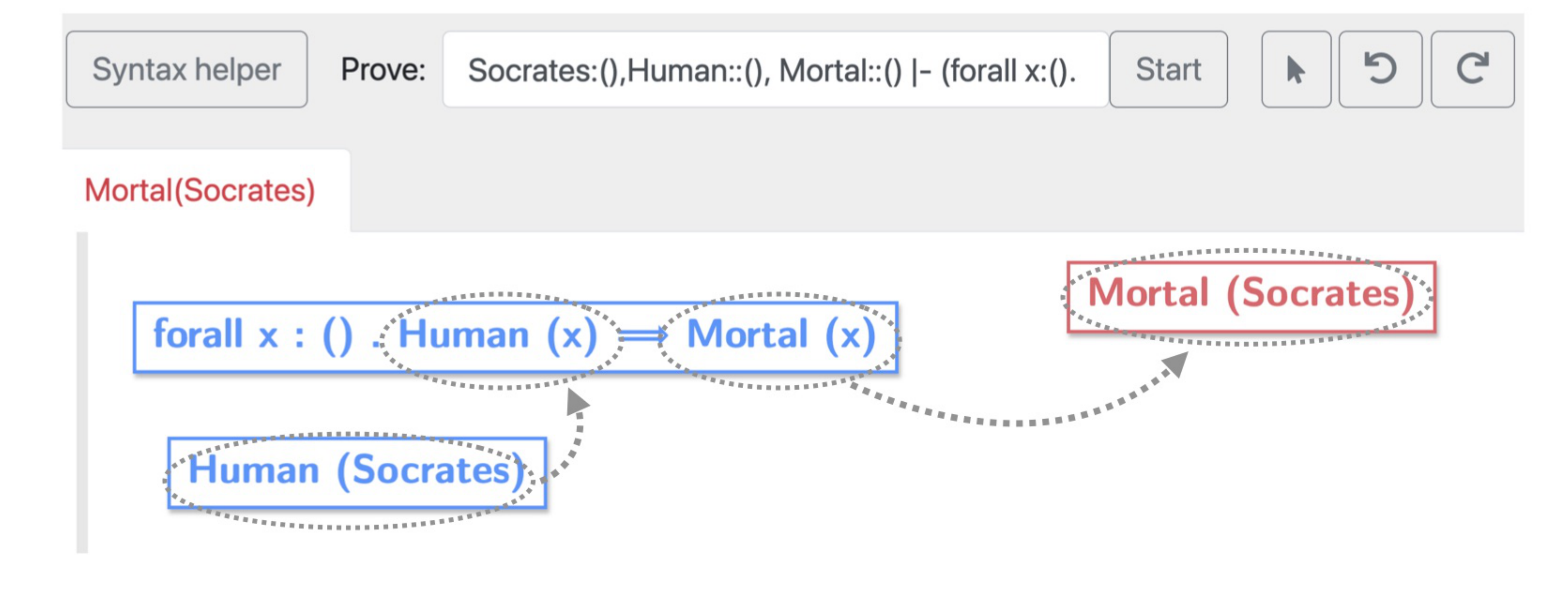}}
 \end{center}
 \caption{A partial screenshot showing a goal in the Actema prototype. The
   conclusion is red on the right, the two hypotheses blue on the
   left. The grey dotted arrows have been added to show the two
   possible actions.}
 \label{fig:aristote}
 \end{figure*}

We will not consider, on paper, the details of variable renaming in
substitutions, implicitly applying the so-called Barendregt
convention, that bound and free variables are distinct and that a
variable is bound at most once.

Extending this work to simple extensions of FOL, like multi-sorted
predicate calculus is straightforward (and actually done in the
prototype).  Some more interesting points may show up when considering
how to apply this work to more complex formalisms like type
theories. We will not explore these questions here.

Another interesting and promising question is how our approach extends
to classical logic(s), that is multi-conclusion classical sequents. In
this text we only give a few hints on this topic.

\section{A First Example}\label{sec:aristote}

\subsection{Layout}
One advantage of the proof-by-actions paradigm, is that it allows a very lean
visual layout of the proof state. There is no need to name hypotheses. In the
prototype we also dispense with a text buffer, since proofs are solely built
through graphical actions.

Figure~\ref{fig:aristote} shows the layout of the system using the
ancient example of Aristotle. A goal appears as a set of {\em items}
whose nature is defined by their respective colors\footnote{We are
  well aware that, in later implementations, this color-based
  distinction ought to be complemented by some other visual
  distinction, at least for users with impaired color vision. But in
  the present description we stick to the red/blue denomination, as it
  is conveniently concise.}:
\begin{itemize}
\item A {\em red item} which is the proposition to be proved, that is the
 {\em conclusion},
\item {\em blue items}, which are the local {\em hypotheses}.
\end{itemize}

{\em The items are what the user can act upon: either by {\em clicking} on
them, or by {\em moving} them.}

Finally, note that each goal is displayed on a tab.

\subsection{Two Kinds of Actions}
In this example, there are two possible actions.

\begin{itemize}
\item A first one is to bring together, by drag-and-drop, the red conclusion $\mortal(\socrates)$ with the succedant of the first hypothesis $\mortal(x)$. This will transform the goal by changing the conclusion to $\human(\socrates)$.
\item A second possibility is to combine the two hypotheses; more precisely to bring together the item $\human(\socrates)$ with the premise $\human(x)$ of the first hypothesis. This will yield a new hypothesis $\mortal(\socrates)$.
\end{itemize}

The first case is what we call a {\em backward step} where the conclusion is
modified by using a hypothesis. The second case is a {\em forward step} where
two known facts are combined to deduce a new fact, that is an additional blue
item.

In both cases, the proof can then be finished invoking the logical
axiom rule. In practice this means bringing together the blue
hypothesis $\human(\socrates)$ (resp. the new blue fact
$\mortal(\socrates)$) with the red goal.

\subsection{Modeling the Mechanism}

A backward step involves a hypothesis, here
$\forall x.\-\human(x) \allowbreak\implies \mortal(x)$ and the
conclusion, here $\mortal(\socrates)$.  Furthermore, the action
actually links together two {\em subterms} of each of these items;
this is written by squaring these subterms. The symbol $\vdash$, used
as an operator, is meant to describe the result of the
interaction. Internally, the behavior of this operator is defined by a
set of rewrite rules given in figures~\ref{fig:DISL} and
\ref{fig:DISL-U}. Here is the sequence of rewrites corresponding to
the example\footnote{Note that $\vdash$ has lower precedence than all
  logical connectives.}: \renewcommand{\arraystretch}{1.1}
$$\begin{array}{lll}
    &  \forall x.\human(x)\implies \select{\mortal(x)} \back \select{\mortal(\socrates)}&\\
    \rew &
           \human(\socrates)\implies \select{\mortal(\socrates)}
           \back \select{\mortal(\socrates)}&
                                               \mathsf{L\forall i}\\
    \rew &
           \human(\socrates)\wedge(\select{\mortal(\socrates)}
           \back \select{\mortal(\socrates)}) &
                                                 \mathsf{L\!\!\lto_2}\\
    \rew &  \human(\socrates)\wedge \top &
                                           \mathsf{id}\\
    \rew & \human(\socrates)&\mathsf{neur}\\
  \end{array}$$

Notice that:
\begin{itemize}
\item   These elementary rewrites are not visible for the
  user. What she/he sees is the final result of the action, that is
  the last expression of the rewrite sequence.
\item The definitions of the rewrite rules in figures~\ref{fig:DISL} and
  \ref{fig:DISL-U} do not involve squared subterms. The information of which
  subterms are squared is only used by the system to decide which rules to
  apply in which order.
\end{itemize}

In general, the action solves the goal when the interaction ends with
the trivially true proposition $\top$. The base case being the action
corresponding to the axiom/identity rule \rnmsf{id}: $A\back A \rew\top$.

A forward step, on the other hand, involves two (subterms of two)
hypotheses. The interaction operator between two hypotheses is written
$*$. In the example above, the detail of the interaction is:
$$
  \begin{array}{lll}
    &  \forall x .\select{\human(x)}\implies\mortal(x) \forw \select{\human(\socrates)}&\\
    \rew & \select{\human(\socrates)}\implies\mortal(\socrates) \forw \select{\human(\socrates)}& \mathsf{F \forall i}\\
    \rew & (\select{\human(\socrates)}\back \select{\human(\socrates)} )\implies\mortal(\socrates) &\mathsf{F\!\!\lto_1}\\
    \rew & \top \implies\mortal(\socrates) &\mathsf{id}\\
    \rew & \mortal(\socrates) & \mathsf{neul}\\
  \end{array}
$$

The final result is the new hypothesis.

We come back to the study of the rewrite rules of $\vdash$ and $*$
further down.


\section{Proof Steps through Clicks}\label{sec:clicks}
Drag-and-drop actions involve two items. Some proof steps involve only
one item; they can be associated to the action of clicking on this
item. The general scheme is that clicking on a connective or quantifier
allows to ``break'' or destruct this connective. The results of clicks
are not very surprising, but this feature is necessary to complement
drag-and-drop actions.
\begin{itemize}
\item Clicking on a blue conjunction $A\wedge B$ transforms the
    item into two separate blue items $A$ and $B$.
\item Clicking on a red conjunction $A\wedge B$ splits the goal into
    two subgoals, whose conclusions are respectively $A$ and $B$.
\item Clicking on a blue disjunction $A\vee B$ splits the goal into two subgoals
    of same conclusion, with $A$ (resp. $B$) added as a new hypothesis.
\item Clicking on the left (resp. right)-hand subterm of a red
      disjunction $A\vee B$ replaces this red conclusion by $A$
      (resp. $B$).
\item Clicking on a red implication $A\implies B$ breaks it into a
      new red conclusion $B$ and a new blue hypothesis $A$.
\item Clicking on a red universal quantifier $\forall x.A$ introduces
  a new object $x$ and the conclusion becomes $A$.
\item Clicking on a blue existential $\exists x.A$ introduces a new
  object $x$ together with a blue hypothesis $A$.
\item Clicking on a red equality $t = t$ solves the goal immediately.
\end{itemize}

One can see that these actions correspond essentially to the
introduction rules of the head connective for the conclusion, and the
elimination rule for the hypotheses.

It is possible to associate some more complex effects to click actions performed
on locations deeper under connectives. This is the essence of proof-by-pointing,
and~\cite{PbP} provides ample description. Since we here focus on drag-and-drop
actions, we do not detail further more advanced PbP features. However we stress
that these features are essentially compatible with what we describe
in this work.

\subsubsection*{Adding New Items}

Often in the course of a proof, one will want to add new items: either a new
conjecture (blue item), or a new object (green item) that would be helpful to
solve the current goal. These can be done respectively with the blue
\texttt{+hyp} and the green \texttt{+expr} buttons, which appear in the
screenshot of figure \ref{fig:edukera}. When clicked, they prompt the user
for the statement of the conjecture, or the name and expression defining the
object. The \texttt{+hyp} button will also create a new subgoal requiring to
prove the conjecture within the current context.

This mechanism and the syntax are for now very crude. The design of
possible smoother tools is an important issue but left for future
work\footnote{For instance~\cite{omar-filling-2021} deals with a
  similar problem in the context of functional programming.}.


\section{A Simple Example Involving Equality}\label{sec:equality}

In most interactive theorem provers, the most basic \emph{rewrite}
tactic allows the use of equality hypotheses, that is known equations
of the form $t = u$, in order to replace some occurrences of $t$ by
$u$ (or symmetrically, occurrences of $u$ by $t$). This substitution
can be performed in the conclusion or in hypotheses. Specifying the
occurrences to be replaced with textual commands can be quite
tedious, since it involves either dealing with some form of naming/numbering,
or writing manually patterns which duplicate parts of the structure of terms.

\begin{figure*}
 \begin{center}
  \fbox{\includegraphics[width=.8\textwidth]{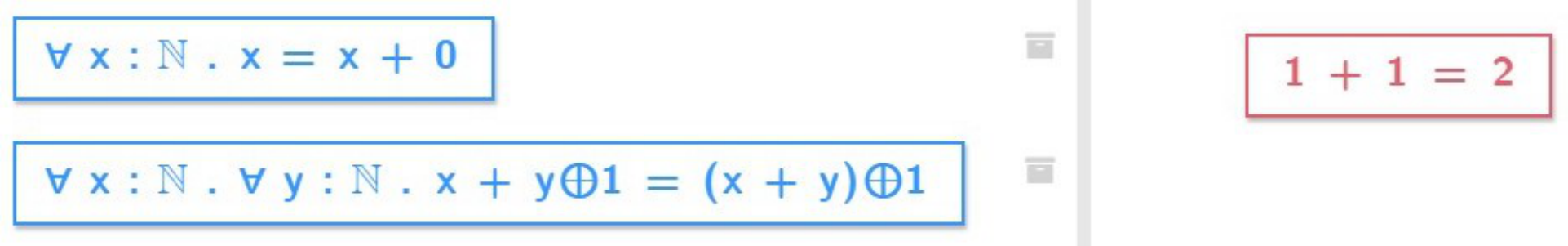}}
 \end{center}
 \caption{Proving $1 + 1 = 2$ in Peano arithmetic.}
 \label{fig:oneplusone}
\end{figure*}

In our setting we can provide this replacement operation through
drag-and-drop. The user points at the occurrence(s) of $t$ to be
replaced, and then brings them to the corresponding side of the
equality.

Figure \ref{fig:oneplusone} shows a very elementary example which is
proving $1+1=2$ in the setting of Peano arithmetic.  For any number
$n$, we write $\suc{n}$ to denote the application of the successor
function to $n$; closed terms are directly written in decimal
notation. The proof goes as follows\footnote{We use the symbol
  $\syneq$ to denote syntactic equality of two expressions modulo
  \emph{pretty-printing}, e.g. decimal notation.}:
\begin{itemize}
  \item We link the left-hand side $x + \suc{y}$ of the second addition axiom with $1 + 1$ in the conclusion, which has the effect of rewriting $1 + 1$ into $\suc{(1 + 0)}$:
    $$
      \begin{array}{lll}
        & \forall x. \forall y. \select{x + \suc{y}} = \suc{(x + y)} \back \select{1 + 1} = 2 & \\
        \rew & \forall y. \select{1 + \suc{y}} = \suc{(1 + y)} \back \select{1 + 1} = 2 &\mathsf{L\forall i} \\
        \rew & \select{1 + \suc{0}} = \suc{(1 + 0)} \back \select{1 + 1} = 2 &\mathsf{L\forall i} \\
        \syneq & \select{1 + 1} = \suc{(1 + 0)} \back \select{1 + 1} = 2 & \\
        \rew & \suc{(1 + 0)} = 2 &\mathsf{L\!\!=_1}
      \end{array}
    $$
  \item We link the right-hand side $x + 0$ of the first addition axiom with $1 + 0$ in the conclusion, which rewrites $1 + 0$ into $1$:
    $$
      \begin{array}{lll}
        & \forall x. x = \select{x + 0} \back \suc{(\select{1 + 0})} = 2 & \\
        \rew & 1 = \select{1 + 0} \back \suc{(\select{1 + 0})} = 2 &\mathsf{L\forall i} \\
        \rew & \suc{1} = 2 &\mathsf{L\!\!=_2} \\
        \syneq & 2 = 2 &
      \end{array}
    $$
\end{itemize}

We end up with the conclusion $2 = 2$, which is provable by a simple click.
Notice how the orientation of the two rewritings is determined by which side of
the equality is selected. Also, in this case, the rewritings
correspond to backward proof steps, because the rewriting is performed
in the conclusion. Similar rules (\rnmsf{F\!\!=_1} and \rnmsf{F\!\!=_2}) are
used to perform rewritings in hypotheses.

\section{Drag-and-Dropping through Connectives}
We mentioned in section \ref{sec:clicks} that it is possible to destruct logical
connectives through click actions. In many cases however, this will not be
necessary: because a drag-and-drop involves subterms of the items involved, one
can often directly use (resp. act on) the part of the hypothesis (resp.
conclusion) which is of interest.

\subsection{Conjunction and Disjunction}
The conjunction is an easy to explain case. A hypothesis of the form
$A\wedge B$ can be used directly both as evidence for $A$ and as evidence
for $B$. This is modeled by the rules \rnmsf{L\land_1} and
\rnmsf{L\land_2}. A very simple action is thus:
$$
\begin{array}{llll}
  \select{A}\wedge B \back \select{A} & \rew& \select{A} \back
  \select{A} \hbox to 1cm {\hfil}&\mathsf{L \land_1}\\
                                       & \rew &\top & \mathsf{id}
\end{array}
$$

On the other hand, considering a conjunctive goal $A\wedge B$, one can
simplify or solve one of the branches by a DnD action. This involves
rules \rnmsf{R \land_1} and
\rnmsf{R \land_2}. For instance:
$$
\begin{array}{llll}
  \select{A} \back \select{A}\wedge B &
                                         \rew& \select{A} \back
                                         \select{A}\wedge B &\mathsf{R \land_1}\\
                                       & \rew& \top \wedge B  & \mathsf{id}\\
  &\rew& B&\mathsf{neul}
\end{array}
$$

Red disjunctions work similarly to conjunctive goals, except that solving one
branch will solve the entire goal. A nice consequence of this, which is hard to
simulate with textual tactics, is that one can just simplify one branch of a
disjunction without comitting to it:
$$
\begin{array}{llll}
  \select{A} \back (B \wedge \select{A}) \vee C
    & \rew & (\select{A} \back B \wedge \select{A}) \vee C &\mathsf{R\lor_1}\\
    & \rew & (B \wedge (\select{A} \back \select{A})) \vee C &\mathsf{R\land_2}\\
    & \rew & (B \wedge \top) \vee C &\mathsf{id}\\
    & \rew & B \vee C &\mathsf{neur}
\end{array}
$$

Disjunctive hypotheses also have a backward behavior defined by the rules
\rnmsf{L\lor_1} and \rnmsf{L\lor_2}, although in most cases one will prefer the
usual subgoal semantics associated with click actions. More interesting is
their forward behavior with the rules \rnmsf{F\lor_1} and \rnmsf{F\lor_2}, in
particular when they interact with negated hypotheses. For instance:
$$
\begin{array}{llll}
  \select{A} \lor B \forw \neg \select{A}
    & \rew & (\select{A} \forw \neg\select{A}) \lor B &\mathsf{F\lor_1}\\
    & \rew & \neg (\select{A} \back \select{A}) \lor B &\mathsf{F\!\!\lto\!\!_1}\\
    & \rew & \neg \top \lor B &\mathsf{id}\\
    & \rew & \bot \lor B &\mathsf{neul}\\
    & \rew & B &\mathsf{neul}\\
\end{array}
$$

We have noticed that on some examples, such actions could provide a
significant speed-up with respect to traditional textual command
provers. We give a more concrete example in section~\ref{sec:edukera}.

Notice that we used rules associated with implication, since negation can be
defined by $\neg A \triangleq A \lto \bot$.

\subsection{Implication}\label{sec:implication}
The implication connective is crucial, because it is not monotone. More
precisely, the roles of hypotheses and conclusions are reversed on the
left of an implication. We start with some very basic examples
for the various elementary cases.

Using the right hand part of a hypothesis $A\implies B$ turns a 
conclusion $B$ into $A$. 
$$
\begin{array}{llll}
  A\implies \select{B} \back \select{B} &\rhd& A\wedge (\select{B}
                                                \back
                                                \select{B})&\mathsf{L
                                                             \!\!\implies_2}\\
                                         &\rhd&A\wedge\top & \mathsf{id}\\
    &\rhd& A&\mathsf{neul}
\end{array}
$$

This can also be done under conjunctions and/or disjunctions:
$$
  A\implies \select{B} \back C \wedge (D\vee \select{B}) ~~ \mrew ~~ C \wedge (D\vee A)
$$

An interesting point is what happens when using implications with
several premisses. The curried and uncurried versions of the
implication will behave exactly the same way:

$$
  A\implies B \implies \select{C} \back D\vee \select{C}
  ~~~\/ ~\mrew~~~  D\vee (A\wedge B)
$$
and:
$$
  A\wedge B \implies \select{C} \back D\vee \select{C}
  ~~~ \/~ ~~\/~~\mrew~~~  D\vee (A\wedge B)
$$

As we have seen in Aristotle's example (section \ref{sec:aristote}), blue
implications can also be used in forward steps, where another
hypothesis matches one of their premisses.

A first nice feature is the ability to strengthen a hypothesis by
providing evidence for any of its premises:
$$
B\implies \select{A}\implies C \forw \select {A}  ~~\/~\/~\mrew~\/~~~~~
B\implies C$$
and again the same can be done for the uncurryfied version:
$$
B\wedge \select{A}\implies C \forw \select {A}  ~\/~\/~\mrew~\/~~~~~
B\implies C.$$

The two aspects of the implication can be combined:

$$
B\implies \select{A}\implies C \forw D\implies\select {A}  ~\/~\/~\mrew~\/~~~~~
B\implies D\implies C$$
or:
$$
B\wedge \select{A}\implies C \forw D\implies\select {A}  ~\/~\/~\mrew~\/~~~~~
B\wedge D \implies C.$$

Note that there is almost no difference in the way one uses different
versions of a hypothesis $A\implies B\implies C$, $A\wedge B\implies
C$, but also $B\implies A\implies C$, in forward as well as in
backward steps\footnote{When viewed as types through the Curry-Howard
isomorphism, $A\implies B\implies C$, $A\wedge B\implies C$, $B\wedge
A\implies C$ and $B\implies A\implies C$ are {\em isomorphic types};
and Roberto di Cosmo~\cite{ISOSBook} has also precisely underlined
that type isomorphisms should help to free the programmer from
arbitrary syntactical choices.}. This underlines, we hope, that our
proposal makes the proof construction process much less dependent on
arbitrary syntactical details, like the order of hypotheses or whether
they come in curryfied form or not.

Also, the rules for implication combined with the rules for equality
\rnmsf{L\!\!=_i} or \rnmsf{F\!\!=_i} naturally give access to {\em
  conditional rewriting}; we detail this in combination with
quantifiers in the next section.


As for red implications, they also have a backward semantics with the rules
\rnmsf{R\!\!\lto\!\!_1} and \rnmsf{R\!\!\lto\!\!_2}, but most of the time one
will want to destruct them immediately by click. An exception could be if one
wants to simplify some part of an implicative, inductive goal before starting
the induction.

\subsection{Quantifiers}
As the first example of this paper shows, the drag-and-drop actions work through
quantifiers and can trigger instantiations of quantified variables. This is made
possible by the rules \rnmsf{L\forall i} and \rnmsf{F\forall i}, which allow the
instantiation of a variable universally quantified in a hypothesis.

Symmetrically, a variable quantified existentially in a conclusion can
also be instantiated. For instance:

$$\begin{array}{llll}
    \select{A(t)}\back \exists x.\select{A(x)}&\rhd&
                                                      \select{A(t)}\back\select{A(t)}&\mathsf{L}\forall\mathsf{i}
    \\
                                               &\rhd& \top&\mathsf{id}
  \end{array}
  $$

An interesting feature is the possibility to modify propositions under
quantifiers. Consider the following possible goal:
$$\forall a.\exists b. A(f(a)+g(b))$$
where $A$, $f$ and $g$ can be complex expressions. Suppose we have a
lemma allowing us to prove:
$$\forall a.\exists b. A(g(b)+ f(a)).$$
Switching from one formulation to the other, involves one use of the
commutativity property $\forall x.\forall y. x+y=y+x$.
In our setting, the equality can be used under quantifiers in one single action:
$$
\begin{array}{ll}
  &\forall x.\forall y. \select{x+y}=y+x \back \forall a.\exists b. A(\select{f(a)+g(b)})\\
\rhd^* & \forall a.\exists b. A(g(b)+f(a))
\end{array}$$

Note also that it is possible to instantiate only some of the universally
quantified variables in the items involved. In general, a universally
quantified variable can be instantiated when the quantifier is in a
negative position; for instance:
$$
\begin{array}{rcl}
 \forall x.\forall
 y. \select{P(y)}\implies R(x,y)\forw \select{P(a)} ~\/~\mrew
~\/~ \forall x. R(x,a)
\end{array}
$$
This last example illustrates how partial instantiation abstracts away the order
in which quantifiers are declared, very much like the partial application
presented in section \ref{sec:implication}.

Again, in some cases, only some existential quantifiers may be
instantiated following a linkage:
$$
\begin{array}{rcl}
\select{P(a)}\back \exists x.\exists y. \select{P(y)}\wedge R(x,y)
 ~\/~\mrew
~\/~
    \exists x. R(x,a)
\end{array}
$$

When using an existential assumption, one can either destruct it
through a click, or use or transform it through a DnD; for instance:
$$
\exists x.\select{P(x)}\forw \forall y.\select{P(y)}\implies
Q(y)  ~\/~\mrew
~\/~\exists x.Q(x)
$$

\subsection{Dependency between Variables}\label{sec:acyclicity}
Some more advanced examples yield simultaneous instantiations of
existentially and universally quantified variables. In such cases, the
system needs to check some dependency conditions. For instance, the
following linkage is valid and solves the goal through one action:
$$
\begin{array}{lll}
  &\exists y. \forall x. \select{R(x,y)}\back \forall x'.\exists
    y'. \select{R(x',y')}& \\
  \rhd& \forall y.(\forall x. \select{R(x,y)}\back \forall x'.\exists
        y'. \select{R(x',y')} ) \mbox{\hbox to 12pt{\hfill}}& \mathsf{L}\exists\mathsf{s}\\
  \rhd &\forall y. \forall x'. (\forall x. \select{R(x,y)}\back \exists
  y'. \select{R(x',y')} ) ~~~~~~~~~~~~&\mathsf{R}\forall\mathsf{s}\\
  \rhd &  \forall y. \forall x'. (\forall
         x. \select{R(x,y)}\back\select{R(x',y)} )&\mathsf{R}\exists\mathsf{i}\\
  \rhd&   \forall y. \forall
           x'. (\select{R(x',y)}\back\select{R(x',y)} )&\mathsf{L}\forall\mathsf{i}\\
   \rhd  &  \forall y. \forall
           x'. \top & \mathsf{id}\\
\rhd^*& \top
\end{array}
$$

But the contraposed situation is not provable; the system will refuse
the following linkage:
$$
\begin{array}{rcl}
  \forall x. \exists y. \select{R(x,y)}\back \exists y'.\forall x'. \select{R(x',y')}
\end{array}
$$
Indeed, there is no reduction path starting from this linkage ending with the
{\sf id} rule. This can be detected by the system because the unification of
$R(x,y)$ and $R(x',y')$ here results in a cycle in the instantiations of
variables\footnote{Also notice that this example requires to use full
(first-order) unification, not only matching.}. The system thus refuses this
action.

\subsection{Conditional Rewriting}
The example given in section~\ref{sec:equality}, although very simple,
already combines the rules for equality and for quantifiers. When also
using implication, one obtains naturally some form of conditional
rewriting. To take another simple example, suppose we have a
hypothesis of the form:
$$\forall x. x\neq 0 \implies f(x) = g(x)$$

We can use this hypothesis for replacing a subterm $f(t)$ by $g(t)$,
which will generate a side-condition $t\neq 0$:
$$
\begin{array}{lll}
  &\forall x. x\neq 0 \implies \select{f(x)} = g(x) \back A(\select{f(t)}) &\\
  \rew & t\neq 0 \implies \select{f(t)} = g(t) \back A(\select{f(t)}) &\mathsf{L \forall i}\\
  \rew & t\neq 0 \wedge (\select{f(t)} = g(t) \back A(\select{f(t)})) & \mathsf{L\!\!\lto_2}\\
  \rew &  t\neq 0 \wedge A(g(t)) &\mathsf{L\!\!=_1} \\
\end{array}$$

One could similarly do such a rewrite in a hypothesis. Furthermore,
the conditional rewrite can also be performed under quantifiers; for instance:
$$
\begin{array}{lll}
  &\forall x. x\neq 0 \implies \select{f(x)} = g(x) \back \exists y . A(\select{f(y)})
  &\mathsf{R \exists s}\\
  \rew &\exists y. ( \forall x. x\neq 0 \implies \select{f(x)} = g(x) \back A(\select{f(y)})) &\mathsf{L \forall i}\\
  \rew & \exists y. ( y\neq 0 \wedge ( \select{f(y)} = g(y) \back A(\select{f(y)}) )) &\mathsf{L\!\!\lto_2} \\
  \rew & \exists y .( y\neq 0 \wedge A(g(t))) &\mathsf{L\!\!=_1} \\
\end{array}$$

\section{Correctness}\label{sec:correctness}

All examples up to now followed the scheme for DnD actions sketched in section
\ref{sec:aristote}:
\begin{itemize}
  \item Given a blue item $A$ and a red item $B$, backward proof steps produce a
  new conclusion $C$ by applying a sequence of rewrite rules $A \vdash B \mrew
  C$.
  \item Given two blue items $A$ and $B$, forward proof steps produce a new
  hypothesis $C$ by applying a sequence of rewrite rules $A \ast B \mrew C$.
\end{itemize}

Thus for such actions to be logically correct, we have to make sure that our
rewrite system satisfies the following property:

\begin{property}[Correctness]\label{prop:correctness}
  \phantom{a}
  \begin{itemize}
    \item If $A \vdash B \mrew C$, then $A, C \vdash B$ is provable.
    \item If $A \ast B \mrew C$, then $A, B \vdash C$ is provable.
  \end{itemize}
\end{property}

We first need a few definitions to handle the fact that rewrite rules apply at
any depth inside formulas:

\begin{definition}[Context]\label{def:context}
  A context, written $A\hole$, is a proposition containing exactly
  one occurrence of a specific propositional variable $\hole$ which is
  not used elsewhere.

  Given another proposition $B$, we write $A\select{B}$ for the
  proposition obtained by replacing $\square$ in $A\square$ by $B$. Note
  that this replacement is not a substitution because it allows variable
  capture. For instance $\forall x.\select{P(x)}$ is the proposition
  $\forall x.P(x)$.
\end{definition}

\begin{definition}[Path]\label{def:path}
A path is a proposition where one subformula has been
selected. Formally, a path is a pair $(A\square, B)$ formed by one
context and one proposition:
\begin{itemize}
\item $A\square$ is called the {\em context} of the path,
\item $B$ is called the {\em selection} of the path.
\end{itemize}

The path $(A\square, B)$ can be viewed as the proposition
$A\select{B}$.  For readability, we will generally also write
$A\select{B}$ for the path $(A\square, B)$.
\end{definition}

\begin{definition}[Inversions]
  Given a context $A\hole$, the number of inversions in $A\hole$,
  written $\inv(A\hole)$, is the number of subterms of $A\hole$
  which are of the form $C\square\implies D$; that is the number of
  times the hole is on the left-hand side of an implication.
  \end{definition}
\noindent  For instance:
  $$
\begin{array}{rcl}
    \inv(D\wedge \hole)&=& 0\\
    \inv((D\wedge \hole)\implies E)&=&1\\
    \inv((\hole\implies C)\implies D) &=& 2
\end{array}
$$

\begin{definition}[Polarity of a context]\label{def:polarity}
We will write $A^+\hole$ to specify that a context is {\em positive},
meaning that $\inv(A^+\square)$ is even. Symmetrically, $A^-\hole$ will be
used for {\em negative} contexts, meaning that $\inv(A^-\hole)$ is odd.
\end{definition}
The following simple covariance and contravariance property will be used
extensively later on:
\begin{property}\label{prop:cov}
  If $\, \Gamma, A\vdash B$ is provable, then so are $\, \Gamma, C^+\select{A}\vdash C^+\select{B}$
  and $\, \Gamma, D^-\select{B}\vdash D^-\select{A}$.
\end{property}

For each rule, interpreting $\vdash$ as $\lto$ and $\ast$ as $\land$ in the
right-hand side is enough to show that the rule satisfies property
\ref{prop:correctness} locally. But for rewritings taking place at occurrences
deeper inside a proposition, we need to consider the polarity of their context.
Using the notation of definition~\ref{def:polarity}, we can state:
\begin{lemma}\label{lemma:rules-valid-in-context}
  \phantom{a}
  \begin{itemize}
    \item If $C^+\select{A\vdash B}\rew D$ then $D \vdash C^+\select{A\lto B}$ is provable.
    \item If $C^-\select{A\vdash B} \rew D$ then $C^-\select{A\lto B}\vdash D$ is provable.
    \item If $C^+\select{A * B} \rew D$ then $ C^+\select{A \land B}\vdash D$ is provable.
    \item If $C^-\select{A * B} \rew D$ then $D \vdash C^-\select{A \land B}$ is provable.
  \end{itemize}
\end{lemma}
\begin{remark}
  For some rules, like \rnmsf{R\!\!\lto_1}, the left-hand and
  right-hand propositions are equivalent:
  $$A \lto B \lto C ~~~\Leftrightarrow~~~ A \land B \implies A$$
  Such rules are called {\em invertible} and their names are tagged by
  $*$. This point will be relevant in section~\ref{sec:invert}.
\end{remark}

An easy but important technical point is that rewrite rules preserve the
polarity of contexts around redexes, in the following precise sense:
\begin{property}\label{prop:rules-preserve-polarity}
  If $C\select{A\vdash B} \rew C'\select{A'\vdash B'}$
  (resp. $C\select{A* B} \rew $ $ C' \allowbreak\select{A'* B'}$) then
  $C\square$ and $C'\square$ have the same polarity.

  If $C\select{A\vdash B} \rew C'\select{A'* B'}$ (resp. $C\select{A*
    B} \rew C'\select{A'\vdash B'}$) then $C\square$
  and $C'\square$ have opposite polarities.
\end{property}

Combining lemma \ref{lemma:rules-valid-in-context} and property
\ref{prop:rules-preserve-polarity}, we obtain the central correctness result about
the rewrite rules:
\begin{lemma}\label{lemma:rewriting-valid-in-context}
  \phantom{a}
  \begin{itemize}
    \item If $C^+\select{A\vdash B}\mrew D$ then $D \vdash C^+\select{A\lto B}$ is provable.
    \item If $C^-\select{A\vdash B} \mrew D$ then $C^-\select{A\lto B}\vdash D$ is provable.
    \item If $C^+\select{A * B} \mrew D$ then $ C^+\select{A \land B}\vdash D$ is provable.
    \item If $C^-\select{A * B} \mrew D$ then $D \vdash C^-\select{A \land B}$ is provable.
  \end{itemize}
\end{lemma}

We do not detail the proof here, but it relies crucially on the covariance and
contravariance property \ref{prop:cov}.

Finally, property \ref{prop:correctness} is obtained as the special case where
the rewriting starts in the (positive) empty context.

\section{Linkages}\label{sec:linkages}

In demonstrating correctness, we focused solely on the two \emph{items} involved
in a DnD action. But every DnD action also specifies the \emph{selection} of a
subterm in each item. We call \emph{linkage} the combined data of the two items
together with the selection, since the intent is to \emph{link} the subterms
to make them interact in some way.

\begin{remark}
In this article we only consider linkages between two subterms, but as noted in
section \ref{sec:equality}, rewriting is an example of action that can benefit
from allowing multiple selections\footnote{In fact, a restricted kind of
multi-occurrence rewrite is already available in the current prototype of
Actema: one just needs to enter \emph{selection mode}, by either toggling the
dedicated button, or holding down the \texttt{shift} key.}.
\end{remark}

Each kind of DnD action is mapped in the system to a specific form of linkage,
which is designed to hold all the information necessary for the correct
execution of the action. In this way the system can automatically search for
linkages of a certain form, and propose to the user all well-defined actions
associated to these linkages.

\begin{remark}
  In the future, one can imagine several DnD actions associated to a given
  linkage. In this case, the user could be queried to choose the action to be
  performed (typically with a pop-up menu). However with the actions considered
  in this article, such ambiguities never arise.
\end{remark}

On the ``items axis'', we already distinguished between backward and forward
linkages, written respectively $A \back B$ and $A \forw B$. If the items are
unspecified, we will write $A \link B$.

Using the ``selection axis'', we can specify a further distinction that was
informal up to now: that of \emph{logical} action and \emph{rewrite} action.
\begin{itemize}
  \item \emph{Logical} linkages link two subformulas. Thus they have the form
  $B\select{A} \link C\select{A'}$.
  \item \emph{Rewrite} linkages link one side of an equality with a first-order
  term. Using liberally the notations from definitions \ref{def:context} and
  \ref{def:path}, they thus have the form \\$B\select{\select{t} = u} \link C
  \select{t'}$ (or symmetrically $B\select{u = \select{t}} \link C
  \select{t'}$).
\end{itemize}

In both cases, we impose the following condition:
\begin{condition}[Unification]\label{cond:unif}
  The linked subterms ($A$ and $A'$ or $t$ and $t'$) must be \emph{unifiable}
  with respect to the variables quantified in their contexts $B\hole$, $C\hole$.
  We do not detail all the constraints of this unification problem, but the
  essential idea is that a variable is unifiable if and only if its quantifier
  is \emph{instantiable}. This in turn can be determined by the polarity of the
  context surrounding the quantifier. One also needs to check that the unifier
  does not create circular dependencies between variables.
\end{condition}

There are also additional restrictions on the polarities of contexts.
Assuming we work in a goal $\Gamma\vdash C$:
  
\begin{condition}[Polarity]\label{cond:pol}
  The following conditions are necessary for the linkage $B\select{A}\link
  D\select{A'}$ to be a logical linkage:
  \begin{enumerate}
    \item $B\select{A}\in \Gamma$,
    \item $D\select{A'} \begin{dcases*}
      \equiv C & if ${\link} \text{is} {\back}$ \\
      \in \Gamma & if ${\link} \text{is} {\forw}$
      \end{dcases*}$
    \item $(\inv(B\hole), \inv(D\hole)) \in \begin{dcases*}
      \{(0,0),(1,1),(0,2)\} & if ${\link} \text{is} {\back}$ \\
      \{(0,1),(1,0)\} & if ${\link} \text{is} {\forw}$
    \end{dcases*}$
  \end{enumerate}

  Similarly, the following conditions are necessary for the linkage
  $B\select{\select{t} = u} {\link} D\select{t'}$ to be a rewrite linkage. Either
  $B\select{\select{t} = u}\in \Gamma$, then:
  \begin{enumerate}
    \item $D\select{t'} \begin{dcases*}
      \equiv C & if ${\link} \text{is} {\back}$ \\
      \in \Gamma & if ${\link} \text{is} {\forw}$
      \end{dcases*}$
    \item $B\hole$ is positive
  \end{enumerate}
  or $B\select{\select{t} = u} \equiv C$, then ${\link} \text{is} \back$ and:
  \begin{enumerate}
    \item $D\select{t'} \in \Gamma$
    \item $B\hole$ is negative
  \end{enumerate}
\end{condition}

\begin{remark}
The main reason for limiting the number of inversions in a path to 2
is that we place ourselves in intuitionistic logic. In classical
logic, one could, for instance, imagine the following behavior:
$$(\select{A}\implies B)\implies C\back \select{A}~~\mrew~~ C\implies A$$
But this would not be valid intuitionistically.
\end{remark}

\begin{definition}[Valid linkage]
  We say that a linkage $\mathcal{L}$ is \emph{valid} if it satisfies conditions
  \ref{cond:unif} and \ref{cond:pol}.
\end{definition}

One understands that for logical linkages, condition \ref{cond:pol} guarantees
that there is one positive and one negative occurrence among $A$ and $A'$. For
rewrite linkages, it guarantees that the equality is in negative position.

One can also check that all the examples given up to here were based on valid
linkages.

\section{Describing DnD Actions}\label{sec:action}

We are now equipped to specify how logical and rewrite linkages translate
deterministically to the backward and forward proof steps shown in all examples.

First some remarks can be made about the rewrite rules of figure \ref{fig:DISL}:
\begin{itemize}
\item The set of rewrite rules is obviously non-confluent. 
\item It is also terminating, because the number of connectives or
  quantifiers under $\ast$ or $\vdash$ decreases\footnote{Except for the {\sf
  Fcomm} rule which is just meant to make the $\ast$ connective symmetric;
  formally, the only infinite reduction paths end with an infinite iteration of
  {\sf Fcomm}.}.
\end{itemize}

As for the rules of figure \ref{fig:DISL-U}, they are both terminating
\emph{and} confluent. Indeed they define a function that eliminates redundant
occurrences of the units $\top$ and $\bot$.

Here is a high-level overview of the complete procedure followed to generate a
proof step:
\begin{enumerate}
\item \textbf{Selection:} the user selects two subterms in two items of the current goal; \label{step:selection}
\item \textbf{Linkage:} this either gives rise to a logical linkage $B\select{A}
  \link C\select{A'}$ (resp. a rewrite linkage $B\select{\select{t} = u} \link
  C\select{t'}$), or does not correspond to a known form of linkage. In this case
  the procedure stops here, and the system does not propose any action to the
  user; \label{step:linkage}
\item \textbf{Unification:} the system tries to unify the selected subterms $A$
  and $A'$ (resp. $t$ and $t'$), which either yields a substitution $\sigma$, or
  fails. In this case we stop like in the previous step;
  \label{step:unification}
\item \textbf{Linking:} the system then chooses a rewriting start\-ing from the linkage. Thanks to
  theorem \ref{thm:productivity}, this re\-writing always ends with a proposition of
  the form $D\select{\sigma(A) \vdash \sigma(A')}$ (resp.
  $D\select{\select{\sigma(t)} = u \link C_0\select{\sigma(t')}}$); \label{step:linking}
\item \textbf{Interaction:} thus one can apply the {\rnmsf{id}} rule (resp. an equality rule in
$\{\mathsf{L\!\!=\!\!_1}, \mathsf{L\!\!=\!\!_2}, \mathsf{F\!\!=\!\!_1},
\mathsf{F\!\!=\!\!_2}\}$); \label{step:interaction}
\item \textbf{Unit elimination:} in the case of a logical action, this creates an occurrence of $\top$,
which is eliminated using the rules of figure \ref{fig:DISL-U}; \label{step:unit-elimination}
\item \textbf{Goal modification:} the two previous steps produced a formula $E$.
  In the case of a forward linkage, a hypothesis $E$ is added to the goal; in
  the case of a backward linkage, the goal's conclusion becomes $E$. In both
  cases, the logical correctness is guaranteed by
  property~\ref{prop:correctness}. \label{step:goal-modification}
\end{enumerate}

\begin{figure}
  \fontsize{10}{10.5}\selectfont
    \renewcommand{\arraystretch}{1.25}
  {\sc Backward}\\
  \begin{mathpar}
    \begin{array}{r@{\quad}c@{\quad}lc}
        {A \seq A}&\rew&\top &\mathsf{id}\\
        {t = u \seq A}&\rew&\subst{A}{t}{u} &\mathsf{L\!\!=_1}\\
        {u = t \seq A}&\rew&\subst{A}{t}{u} &\mathsf{L\!\!=_2}\\[1em]

        {(B \land C) \seq A}&\rew&        {B \seq A}&\mathsf{L \land_1}\\
        {(C \land B) \seq A}&\rew&        {B \seq A}&\mathsf{L \land_2}\\
        {A \seq (B \land C)}&\rew&        {(A \seq B) \land C}&\mathsf{R \land_1}\\
        {A \seq (C \land B)}&\rew&        {C \land (A \seq B)}&\mathsf{R \land_2}\\[1em]
        
        {(B \lor C) \seq A}&\rew&        {(B \seq A) \land (C \lto A)}~~~~~~~~~&\mathsf{L \lor_1}\rever\\
        {(C \lor B) \seq A}&\rew&        {(C \lto A) \land (B \seq A)}&\mathsf{L \lor_2}\rever\\
        {A \seq (B \lor C)}&\rew&        {(A \seq B) \lor C}&\mathsf{R \lor_1}\\
        {A \seq (C \lor B)}&\rew&        {C \lor (A \seq B)}&\mathsf{R \lor_2}\\[1em]

        {(C \lto B) \seq A}&\rew&        {C \land (B \seq A)}&\mathsf{L\!\!\lto_2}\\
        {A \seq (B \lto C)}&\rew&        {(A \ast B) \lto C}&\mathsf{R\!\!\lto_1}\rever\\
        {A \seq (C \lto B)}&\rew&        {C \lto (A \seq B)}&\mathsf{R\!\!\lto_2}\rever\\[1em]


        {(\forall x. B) \seq A}&\rew&        {\subst{B}{x}{t} \seq A}&\mathsf{L \forall i}\\
        {(\forall x. B) \seq A}&\rew&        {\exists x. (B \seq A)}&\mathsf{L \forall s}\\
        {A \seq (\forall x. B)}&\rew&        {\forall x. (A \seq B)}&\mathsf{R \forall s}\rever\\[1em]

        {(\exists x. B) \seq A}&\rew&        {\forall x. (B \seq A)}&\mathsf{L \exists s}\rever\\
        {A \seq (\exists x. B)}&\rew&        {A \seq \subst{B}{x}{t}}&\mathsf{R \exists i}\\
        {A \seq (\exists x. B)}&\rew&        {\exists x. (A \seq B)}&\mathsf{R \exists s}\\
    \end{array}
  \end{mathpar}
  ~\\[1em]
  {\sc Forward}\\
  \begin{mathpar}
    \begin{array}{r@{\quad}c@{\quad}lc}
      {A \ast (t = u)} &\rew& {\subst{A}{t}{u}}&\mathsf{F\!\!=_1}\\
      {A \ast (u = t)} &\rew& {\subst{A}{t}{u}}&\mathsf{F\!\!=_2}\\[1em]

        {A \ast (B \land C)} &\rew&   {A \ast B}&\mathsf{F\land_1}\\
        {A \ast (C \land B)}&\rew&        {A \ast B}&
            \mathsf{F \land_2}\\[1em]

        {A \ast (B \lor C)}& \rew&       {(A \ast B) \lor C}
      &
      \mathsf{F \lor_1}\\
        {A \ast (C \lor B)}&\rew&        {C \lor (A \ast B)}&
            \mathsf{F \lor_2}\\[1em]

        {A \ast (B \lto C)}
&\rew&        {(A \seq B) \lto C}
      &\mathsf{F\!\!\lto_1}\\
        {A \ast (C \lto B)}&\rew&        {C \lto (A \ast B)}&
            \mathsf{F\!\!\lto_2}\\[1em]


        {A \ast (\forall x. B)}
&\rew&        {A \ast \subst{B}{x}{t}}
      &
      \mathsf{F \forall i}\\
        {A \ast (\forall x. B)}&\rew&        {\forall x. (A \ast B)}&
            \mathsf{F \forall s}\\[1em]

              {A \ast (\exists x. B)}&\rew&{\exists x. (A \ast B)}&
              \mathsf{F \exists s}\rever\\[1em]
              
          {B \ast A}&\rew&
          {A \ast B}& \mathsf{Fcomm}\\
    \end{array}
    \end{mathpar}
  ~\\[1em]
  In the rules $\{\mathsf{L \forall s}, \mathsf{L \exists s}, \mathsf{R \forall
  s}, \mathsf{R \exists s}, \mathsf{F \forall s}, \mathsf{F \exists s}\}$, $x$
  is not free in $A$.
  \caption{Linking rules}
  \label{fig:DISL}
\end{figure}

\begin{figure*}
  \fontsize{10}{10.5}\selectfont
  {\sc Units}\\
  \begin{mathpar}
    \renewcommand{\arraystretch}{1.2}
    \begin{array}{lr@{\quad}c@{\quad}lc}
   \pair{\circ}{\dagger} \in \{
      \pair{\land}{\top},
      \pair{\lor}{\bot},
      \pair{\lto}{\top}
      \}&
      {\dagger \circ A}&\rew&A&\mathsf{neul}\\
 \pair{\circ}{\dagger} \in \{
      \pair{\land}{\top},
      \pair{\lor}{\bot}
      \}&
      {A \circ \dagger}&\rew&A&\mathsf{neur}\\
   \pair{\circ}{\dagger} \in \{
      \pair{\land}{\bot},
      \pair{\lor}{\top}
      \}&
      {\dagger \circ A}&\rew&\dagger&
       \mathsf{absl}\\
  \pair{\circ}{\dagger} \in \{
      \pair{\land}{\bot},
      \pair{\lor}{\top},
      \pair{\lto}{\top}
      \}&
      {A \circ \dagger}
      &\rew&\dagger& \mathsf{absr}\\
  \pair{\diamond}{\dagger} \in \{
    \pair{\forall}{\top},
    \pair{\exists}{\bot}
    \}&      {\diamond x. \dagger}&\rew&\dagger&    \mathsf{absq}
    \\
&      {\bot \lto A}&\rew&\top&\mathsf{efq} 
      \end{array}
  \end{mathpar}
  \caption{Unit elimination rules}
  \label{fig:DISL-U}
\end{figure*}

\subsection{Productivity}

An important property of the linking step \ref{step:linking} is that there is
always a rewriting sequence that brings together the selected subterms, which
ensures that we can proceed to the interaction step \ref{step:interaction}.





Because the rewrite rules are terminating, the important point is to show that
one can always apply a rule until one reaches an interaction rule on the
linkage. In other words, it is possible to find at least one rule which
preserves conditions \ref{cond:unif} and \ref{cond:pol} on linkages:

\begin{lemma}[Valid Progress]\label{thm:vprogress}
  If a linkage $\mathcal{L}$ is valid, then either:
  \begin{enumerate}
    \item $\mathcal{L} \in \{
        \select{A}\vdash \select{A},\,
        \select{t} = u \vdash A\select{t},\,
        u = \select{t} \vdash A\select{t},\, 
        \select{t} = u \ast A\select{t},\,
        u = \select{t} \ast A\select{t}
      \}$;
    \item or $\mathcal{L} \rew D\select{\mathcal{L'}}$ with
      $\mathcal{L'}$ valid.
  \end{enumerate}
\end{lemma}

The proof is not fundamentally difficult, but understandably verbose. The two
main points are:
\begin{itemize}
\item The rules involving a connective always preserve validity.
\item When one can apply a rule involving a quantifier $\forall x$ (resp.
  $\exists x$), one checks whether the substitution instantiates $x$ or not. In
  the first case one performs the instantiation rule \rnmsf{L\forall_i} or
  \rnmsf{F\forall_i} (resp. \rnmsf{R\exists_i}); in the second case the
  corresponding {\sf s} rule.
\end{itemize}


Then we can state the following \emph{productivity theorem}, which is a direct
consequence of the previous lemma and the fact that the rewrite rules
terminate:

\begin{theorem}[Productivity]\label{thm:productivity}
If $\mathcal{L}$ is a valid linkage, then there
is a sequence of reductions with one of the following forms:
\begin{mathpar}
  \mathcal{L} \mrew D^+\select{\select{A} \back \select{A}} \\
  \mathcal{L} \mrew D\select{\select{t} = u \link A\select{t}} \and
  \mathcal{L} \mrew D\select{u = \select{t} \link A\select{t}}
\end{mathpar}
\end{theorem}



\subsection{Choosing the Best Derivation}\label{sec:invert}

A last point to deal with is non-confluence and in particular choosing
between first simplifying the head connective on the right or the left
of $\ast$ or $\vdash$. For instance in
$\select{A}\vee B \vdash B\vee\select{A}$ one can apply either
\rnmsf{L\vee_1} or \rnmsf{R\vee_2}.

Interestingly, an answer is provided by {\em
  focusing}. It has been noticed by Andreoli~\cite{andreoli1992} that, in
bottom-up proof search, one should apply the invertible logical
rules first. In our framework, this translates into first applying the
invertible rewrite rules (the ones marked by a *).  In the case of the
example above, this means performing \rnmsf{L\vee_1} first, which
leads to the following behavior:
$$\select{A}\vee B \back B\vee\select{A}\mrew B\implies B\vee A.$$
This is indeed the ``right'' choice, since applying \rnmsf{R\vee_2}
first would lead to a dead-end:
$$\select{A}\vee B \back B\vee\select{A}\mrew B\vee(B\implies A).$$

When two invertible rules can be applied, the order is irrelevant. There are
cases where two non-invertible rules can be applied. The vast majority of them
commute in terms of provability, but not necessarily in the shape of the
resulting formula\footnote{In fact the only rules which do not give equivalent
results when commuted are the critical pairs \rnmsf{F\!\!\lor_i /~F\!\!\lto_2}
for $i \in \{1,2\}$, as was noted independently in
\cite{DBLP:conf/cade/Chaudhuri21}.}. Therefore our specification still leaves
room for some choices. Currently, we have a heuristic prioritizing of the rules
that sticks to what is presented in examples. One could also choose to leave the
disambiguation to the user, e.g. by looking at the \emph{orientation} of
drag-and-drops. This is the solution chosen in \cite{Chaudhuri2013} and
\cite{DBLP:conf/cade/Chaudhuri21}.

\section{A More Advanced Example}\label{sec:edukera}
\begin{figure*}
  \fbox{\includegraphics[width=0.8\textwidth]{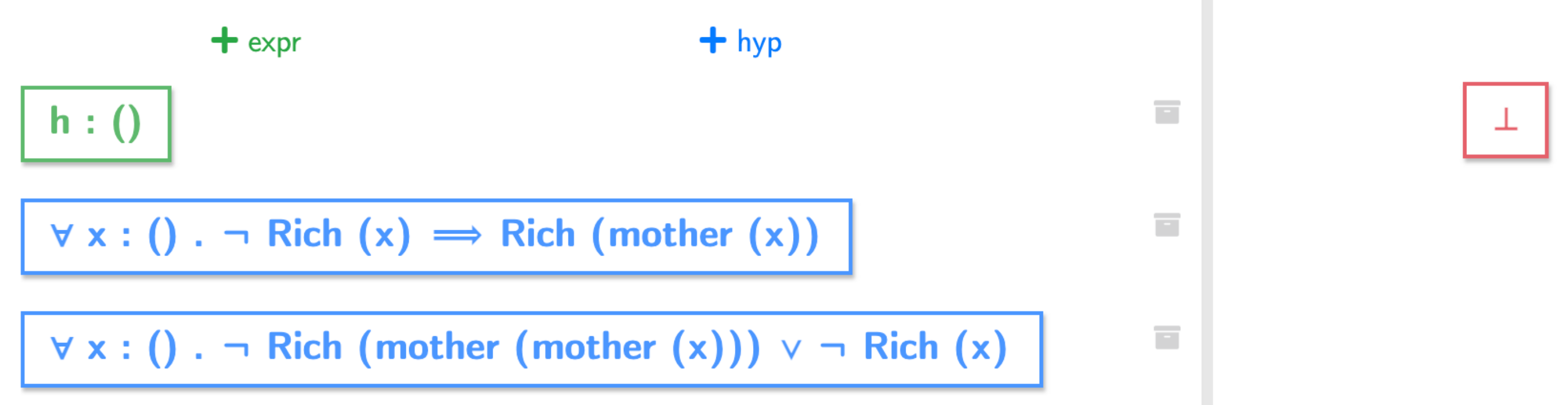}}
\caption{The beginning of an example due to Edukera.}\label{fig:edukera}
\end{figure*}

It is too early to perform a detailed case study comparing our approach
to interactive theorem proving with others -- tactic based,
declarative, {\em etc}\dots~This is due primarily to the fact that
our prototype is not mature enough; it cannot handle lemmas and
implements a limited formalism. However some examples allow to get a
glimpse of specificities and possible advantages of proofs by actions.

One such example is a small logical riddle, which we borrow from a
textual educational system, Edukera~\cite{edukera}. One considers a
population of people, with at least one element $h$, together with a
single function $\mother$ and one predicate $\rich$. The aim is to
show that the two following assumptions are incompatible:
\begin{itemize}
\item[(1)] $\forall x. \neg\rich(x)\vee \neg\rich(\mother(\mother(x)))$,
\item[(2)] $\forall x. \neg\rich(x) \implies \rich(\mother(x)).$
\end{itemize}
The original goal thus corresponds to the illustration of
figure~\ref{fig:edukera}.

It is quite natural to approach this problem in a forward manner, by starting
from the hypotheses to establish new facts. And a first point illustrated by
this example is that DnD actions allow to do this in a smooth and precise
manner. A possible first step is to bring $h$ to the first hypothesis, to obtain
a new fact:

\medskip
$(3) ~\neg\rich(h)\vee \neg\rich(\mother(\mother(h))).$
\medskip

\noindent Double clicking on this new fact yields two cases:
\begin{itemize}
 \item[(4)~] $ \neg\rich(h)$,
 \item[(4')] $ \neg\rich(\mother(\mother(h)))$.
 \end{itemize}
Let us detail how one solves the
second one.

By bringing 
$\neg\rich(\mother(\mother(h)))$ on the premise of $\forall
x. \select{\neg\rich(x)} \implies \rich(\mother(x))$
one obtains

\medskip
$(6) ~\rich(\mother(\mother(\mother(h)))).$
\medskip

The next step is a good example where the DnD is useful. By bringing
this new fact to the right-hand part of

\medskip
$(1)~\forall x. \neg\rich(x)\vee \neg\select{\rich(\mother(\mother(x)))}$
\medskip

\noindent
one immediately obtains a new fact

\medskip
$(7) ~\neg{\rich(\mother(h))}$.
\medskip

\noindent In other proof systems, this last step requires a somewhat intricate
tactic line and/or writing down at least the statement of the new fact.

One can then finish the case by combining $(7)$ and $(2)$ which yields
$\rich(\mother(\mother(h)))$, which contradicts $(4')$. These two last
steps each correspond to a simple DnD.
The other case, $\neg\rich(h)$, is quite similar.

Such a simple example is not sufficient to provide significant
metrics. Note however that once a user has understood the proof, the
riddle is routinely solved in less than a minute in Actema, which
seems out of reach for about any user in a tactic based prover. At
least as important is the fact that the proof can be performed without
typing any text, especially no intermediate statement. 

\section{Related Work}\label{sec:related-work}

\subsubsection*{Subformula Linking}
Although the primary motivation is very practical, it benefitted a lot
from recent proof theory, especially Deep Inference. A key step was
the discovery of the work of Kaustuv
Chaudhuri~\cite{Chaudhuri2013} who had noticed how formula linking in
deep inference could be used for proof construction in linear
logic. His calculus of structures was very important for designing the
rewrite rules which underly our system. In more recent
work~\cite{DBLP:conf/cade/Chaudhuri21} he also deals with
intuitionistic logic.  Interestingly, some ideas like forward proof
steps or the use of colors appeared independently in his and our work.

A difference is that we give the possibility to link first-order terms
in addition to propositions, which is the basis for rewrite
actions. One can imagine to design new kind of transformations in the
future.


\subsubsection*{Window Inference}
We have already mentioned Proof-by-Pointing, which was part of the CtCoq and
Pcoq efforts \cite{amerkad-mathematics-2001} to design a graphical user
interface for the Coq proof assistant. Another contemporary line of work was the
one based on \emph{window inference}, initially pioneered by P.J. Robinson and
J. Staples. In \cite{robinson-formalizing-1993}, window inference is described
as a general proof-theoretical framework, which aims to accomodate for the
pervasive use of \emph{equivalence transformations} throughout mathematics and
computer science.

Window inference has been used both for general-purpose logics like
HOL \cite{grundy-window-1991}, and in more specialized settings like
program refinement \cite{grundy-window-1992}. It naturally lends
itself to integration in a graphical user interface
(\cite{langbacka-tkwinhol-1995}, \cite{goos-tas-2000}), where the user
can \emph{focus} on a subexpression by clicking on it. One is then
presented with a new \emph{graphical} window, holding the selected
expression as well as an extended set of hypotheses exposing
information inferrable from the context of the expression. The user
can pick from a list of valid transformations to be applied to the
expression, before closing the window. This propagates the
transformations to the parent window by replacing the old
subexpression by the new one, without modifying the surrounding
context.

This process is quite reminiscent of the rewriting produced by our DnD
actions.  One key difference is that window inference rules can be
applied stepwise, while we choose to hide the sequence of rules that
justifies a DnD. The window inference approach gives to the user a
precise control of the transformations to be performed and thus could
inspire interesting extensions of our work.


\subsubsection*{Tangible Functional Programming}
We noticed an interesting connection with the work of Conal Elliott on
tangible functional programming \cite{elliott-tangible}. His concept
of \emph{deep application} of $\lambda$-terms seems related to the
notion of subformula linking, when viewing function and product types
as implications and conjunctions through the formulae-as-types
interpretation. He also devised a system of basic combinators which
are composed sequentially to compute the result of a DnD, though it
follows a more complex dynamic than our rewrite rules. Even if the
mapping between proofs and programs is not exact in this case, it
suggests a possible interesting field of application for the
Curry-Howard correspondance, in the realm of graphical
proving/programming environments.

\subsubsection*{Other Gestural Proof Systems}
There are other proof systems which include drag-and-drop features. Two of them
are the KeY Prover \cite{ahrendt-using-2016} and TAS \cite{goos-tas-2000}. TAS
is a window inference system tailored for program refinement, and uses DnD
actions between an expression and a transformation, in order to apply the latter
to the former.
As for the KeY Prover, its usage of DnD overlaps only a very small
portion of usecases that we hinted at in section \ref{sec:edukera}, namely
the instantiation of quantifiers with objects.

We can also mention the recent work of Zhan et
al. \cite{zhan-design-2019}. They share with us the vision of a proof assistant
mainly driven by gestural actions, which requires far less textual inputs from
the user. However, they only consider point-and-click actions, and rely on a
text-heavy presentation at two levels:
\begin{enumerate}
  \item the proof state, which is a structured proof text in the style of Isar~\cite{isar};
  \item the proof commands, which can only be performed through choices in textual menus.
\end{enumerate}


\subsubsection*{Explicit Proof Objects}
Finally let us mention various recent implementations proposing
various ways to construct proofs graphically: Building
Blocks~\cite{buildingblocks}, the Incredible Proof
Machine~\cite{blanchette-visual-2016}, Logitext\footnote{\url{http://logitext.mit.edu/main}} and Click \&
coLLecT~\cite{clickcollect}. But these systems focus more on
explicating the proof object than on making its construction easier.

\section{Conclusion and Perspectives}
This work started as a very practical effort. Discovering and
understanding the links with more theoretically grounded approaches,
and especially deep inference, made us aware that there may be more
proof theoretical depth to this idea than we first thought. But, most
importantly, adapting the logical rules and tools of deep inference to
the practical question we encountered, allowed us to structure our
proposal and to define the ``right'' behavior for the system. We were
able to extend the deep inference approach to the use of
equalities~\ref{sec:equality}, which may be an originality of this
work. It seems imaginable to proceed similarly with other mathematical
relations. 

More generally, we hope that our treatment of equality can be the
start for providing graphical or gestural tools to perform algebraic
transformations of expressions (be there in the conclusion or in
hypotheses). As mentioned above, Window Inference could serve as an
inspiration here. This seems promising to us, since describing such a
transformation is notoriously tedious when using textual commands.

Even a small prototype allowed us to experiment on some non-trivial
examples and to make some first encouraging experiences. In various
cases, like the one described in section~\ref{sec:edukera}, we have
observed shorter or more straightforward proofs than in textual
provers. Another nice point is that some syntactical details, like the
name of proof tactics become irrelevant in the gestural setting. More
generally, we feel that using such a system, one may indeed develop a
good intuition for the behavior of the logical items. But this is
obviously a user interface or user experience question which is too
early to quantify. Also, some novel questions appear when implementing
such a graphical system: what are the good user interface choices, how
to obtain a good look-and-feel, what visual feedback the system should
provide\dots

On the other hand, we should acknowledge that certain styles of proofs,
where a large number of subcases can be immediately solved through the
same short textual tactic sequence, may be less well suited for the
gestural approach (the SSReflect~\cite{SSR} dialect for Coq is very
well suited for such cases).

Among future lines of work, it will be interesting to explore how some
automation fits into this framework. One example is the \emph{point-and-shoot}
paradigm of \cite{PbP}. But the DnD feature could open up new possibilities,
like having the system perform some automated deduction to prove equivalences or
implications between the two squared formulas (which would thus no longer be
required to be strictly equal or unifiable).

Another obvious and important point to be tackled next is to provide a
smooth way to invoke a library of lemmas in a graphical proof. We
believe this could raise some interesting questions.

An also promising line of work is to extend our approach to classical
logic. A point being that the graphical setting could smoothly handle
multiple conclusions with less spurious overhead than text commands.

An important difference with the days of the
pioneering work on proof-by-pointing is that developers can now rely on
powerful and standardized libraries, which make the construction of
user interfaces much faster and easier, giving new room for
experimentation and proposals. But bringing everything together in
simple commands remains a complicated theoretical and development task.

\vspace{0.5em}

\section*{Acknowledgments}

We are grateful to Kaustuv Chaudhuri and Dale Miller for stimulating
discussions, and to Sébastien Najjar of the Dioxygen company for his work on the
front-end of the Actema prototype. Useful comments and references were provided
by anonymous referees.

\nocite{*}
\balance
\bibliography{cpp-article}

\end{document}